\documentclass[12pt]{iopart}
\usepackage{latexsym}
\usepackage{amsfonts}
\usepackage{epsfig}
\usepackage{amssymb}
\usepackage{mathrsfs}
\usepackage{hyperref}
\usepackage{subfigure}

\begin{document}

\title{Effect of hydrogen bonds on protein stability}

\author{Valentino Bianco$^1$, Svilen Iskrov$^{1,2}$, Giancarlo Franzese$^1$}

\address{$^1$Departament de F\'{\i}sica Fonamental, Universitat de Barcelona\\
Diagonal 647, 08028 Barcelona, Spain\\
$^2$\'Ecole Normale Sup\'erieure de Cachan, 
61, avenue du Pr\'esident Wilson
94235 Cachan cedex, France}

\ead{gfranzese@ub.edu}

\begin{abstract}
The mechanism of cold- and pressure-denaturation are matter of debate. Some models
propose that when denaturation occurs more hydrogen bonds between
the molecules of hydration water are formed. Other models identify the
cause in the density fluctuations of surface water, or the
destabilization of hydrophobic contacts because of the displacement of
water molecules inside the protein, as proposed for high
pressures. However, it is clear that water plays a fundamental role in
the process. Here, we review some models that have been proposed to
give insight into this problem. Next we describe a coarse-grained
model of a water monolayer that successfully reproduces the complex
thermodynamics of water and compares well with experiments on proteins
at low hydration level. We introduce its extension for a homopolymer in contact with 
 the water monolayer and study it by  Monte Carlo simulations. 
Our goal is to perform a step in the direction of understanding how
the interplay of cooperativity of water and interfacial hydrogen bonds
affects the protein stability and the unfolding. 
\end{abstract}

\vspace{2pc}

\maketitle

\section{Introduction}

One of the most intriguing challenge in biological physics is the
nature of protein folding-unfolding processes. The temperature range
of stability of proteins is in general small. For example, staphylococcal
nuclease (Snanse--a small protein containing 149 amino-acids) folds 
at low pressure is approximately
between 260~K and 320~K  \cite{ravindra}.

Heat destabilizes proteins. By increasing the bath temperature $T$,
thermal fluctuations increase and disrupt the folded
configurations of proteins. Usually, by decreasing $T$, proteins
crystallize, but surprisingly some proteins unfold at sufficient low
temperature instead of crystallizing \cite{ravindra, pastore,
  privalov, nash, nash2, meersman2, goossens}. 
Cold denaturation seems to be a general phenomenon for proteins,
generally occurring well below $0^{\circ}$C, the freezing point of
water.
In same cases, for example for Snase \cite{ravindra}, the cold
denaturation cannot be directly observed, but experimental data
can be extrapolated to predicted the lower temperature of stability
for the protein. More generally, to make the cold denaturation
observable destabilizing agents can be used.
Interestingly, Pastore et al. \cite{pastore} observed that Yeast
frataxin under physiological conditions undergoes cold denaturation 
below $7^{\circ}$C and remains folded up to $30^{\circ}$C. 
Hence, Yeast frataxin could be an excellent prototype for studying
folding-unfolding  transition, both hot and cold, under accessible
conditions. 

Proteins can unfold also by pressurization.
It has been observed that the increase of pressure
induces the unfolding of protein \cite{meersman, hummer}. 
The pressure-unfolding process can be rationalized by 
considering that the folded structure usually includes cavities.
High pressure can induce elastic response of the protein, deforming
its structure and pushing water molecules inside the cavities. 
The water molecules from inside would 
swell the protein, with consequent loss of protein
functionality \cite{meersman}.
Because is difficult to to separate the protein response
to high hydrostatic pressure from the response of the
aqueous environment, the understanding of 
the problem is still under debate.

\subsection{Thermodynamics of proteins unfolding}

By increasing the thermal energy $k_{B}T$ ($k_{B}$ is the
Boltzmann constant), the protein residues vibrate
faster, accessing new possible configurations, i. e. 
increasing the
the entropy $S$ of the system. This increase leads to
the hot denaturation, in the same way  an increase of $k_{B}T$ leads to the
melting of a crystal. 

The cold denaturation instead, cannot be explained as the effect of an
increase of entropy. By decreasing $T$, the entropy of the system
decreases. This is why we cannot melt a crystal by cooling. Hence,
in the case of proteins there must be a complex mechanism that induces
the cold denaturation. To understand this mechanism is necessary to
introduce the concept of Gibbs free energy $G\equiv H-TS$,  where
$H\equiv U+PV$ is the the enthalpy of the system, $U$ internal energy
of the system, $V$ the volume and $P$ the pressure.

General principles of thermodynamics tell us that at any value of $T$
and $P$ the system minimizes its Gibbs free energy, where 
the system, in our case, is the solution of 
proteins and water. Hence,
the free energy balance must take into account both 
water molecules and protein residues. The experimental fact that
solvated proteins 
unfolds by decreasing $T$ means that at lower $T$ the difference
\begin{equation} \label{var-gibbs}
 \Delta G\equiv G_{u}-G_{f}  
\end{equation}
between the unfolded ($u$) and folded ($f$) states is  
\begin{equation}\label{gibbs}
 \Delta G =
\Delta H-T\Delta S<0,
\end{equation} 
where $\Delta H\equiv H_{u}-H_{f}$ and  $\Delta S\equiv
S_{u}-S_{f}$. 

The total 
variation of the entropy of the system is given by $\Delta S=\Delta
S_{p}+\Delta S_{w}$ where $\Delta S_{p}$ and $\Delta S_{w}$
are the variation of entropy of the protein residues and
water molecules, respectively. 
By unfolding, the protein entropy
increases, $\Delta S_{p}>0$. On the
other hand, the protein contribution to $\Delta H$ is positive, $\Delta H_{p}>0$,
because the enthalpy of the protein increases when the protein unfolds
($H_{p}$ is proportional to the number of contact points of the
protein). Therefore, the protein contribution to $\Delta G$, $\Delta H_{p}-T\Delta
S_{p}$, could be negative or positive depending on the relative
variations and on $T$ and does not guarantees that Eq.~(\ref{gibbs}) is
satisfied. Hence, water contribution to the total balance of Eq.~
(\ref{gibbs}) could be relevant.
To date, is widely shared the idea that the native--folded state is
stabilized by the quasi--ordered network of water molecules hydrating
the non--polar monomers \cite{creighton}.


\subsection{Protein Phase Diagram}

\begin{figure}
\centering
 \includegraphics[scale=0.15]{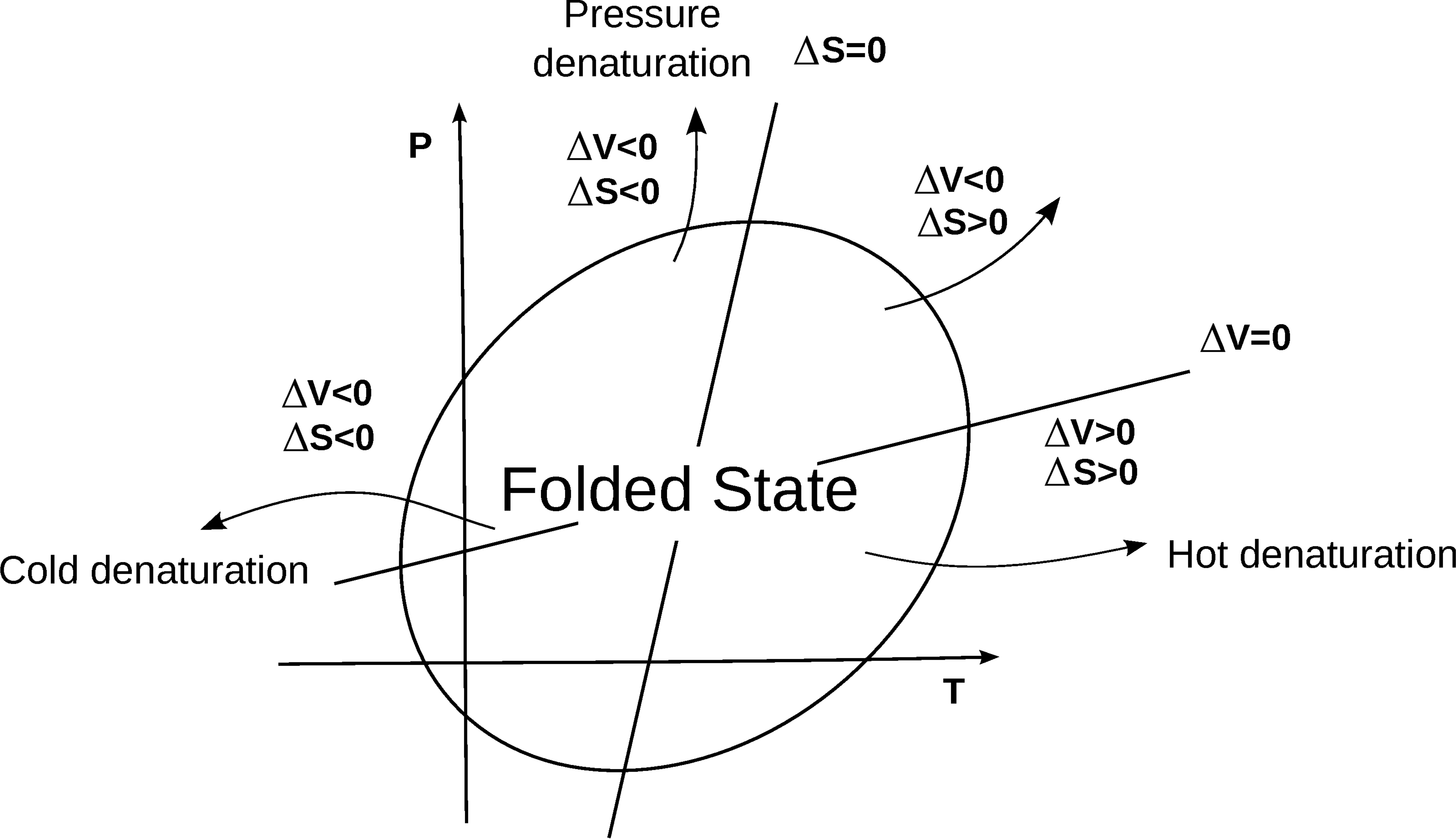}
\caption{Schematic representation of the phase diagram
  of a protein. 
Within the elliptic shape the protein is folded, while it unfolds
by increasing temperature $T$ (hot denaturation), by decreasing $T$
(cold denaturation), by increasing or decreasing pressure $P$
(pressure denaturation). Each folding--unfolding process is characterized
by different variation of entropy $\Delta S$ and variation of  volume
$\Delta V$. The axes of the ellipse are
 loci where $\Delta S=0$ and $\Delta
V=0$ (see text for discussion). Adapted from \cite{meersman}.} 
\label{protein-phase-diagram}
\end{figure}

Experiments are consistent with a protein stability phase
diagram with an elliptic shape \cite{ravindra, pastore, privalov,
  nash,nash2,kunugi} in $P-T$ plane (Fig.~\ref{protein-phase-diagram}). 
Outside the elliptic region the protein unfolds loosing its biological
function. Following Hawley \cite{hawley, smeller}, we can 
calculate $\Delta G$ of the whole system (protein and water) assuming
that a protein can stay in only two distinct states, folded and
unfolded  as in Eq.~(\ref{var-gibbs}).

Because
the internal energy of the system $U$
is a state variable, we can
express its infinitesimal variation $dU$ as a 
function of two thermodynamical quantities. 
If we assume that the unfolding process can be described by
infinitesimal quasi-static transformations, applying 
the First Law of
Thermodynamics, we get
\begin{equation}
 dU=\delta Q - \delta W  , 
\end{equation} 
where $\delta Q$ and $\delta W$ are 
the infinitesimal heat absorbed and the infinitesimal work done by
system, respectively,  along the
generic transformation. Since at constant $T$ and $P$ is $\delta
Q=TdS$ and $\delta W=PdV$, we can 
express the internal energy variation, for a constant number of
particles $N$, as 
\begin{equation}
 dU=TdS-PdV\equiv dU(S,V)  .
\end{equation} 
Differentiating $H=U+PV$
we get
\begin{equation}\label{diff-entalpia}
 dH=dU+PdV+VdP=TdS+VdP\equiv dH(S,P),
\end{equation} 
and differentiating $G=H-TS$, we finally get 
\begin{equation}
 dG=dH-TdS-SdT=-SdT+VdP \equiv dG(T,P) .
\end{equation} 
Hence, it is
\begin{equation} \label{dG}
 d\Delta G=-\Delta SdT+\Delta VdP
\end{equation}
with $\Delta S\equiv S_{u}-S_{f}$ and $\Delta V\equiv V_{u}-V_{f}$.
By expanding 
$\Delta S$ and $\Delta P$ to the first order around 
$\Delta S_{0}$ and
$\Delta V_{0}$,
changes  at $T_{0}$ and $P_{0}$,
we get
\begin{equation}\label{deltaS}
 \Delta S=\Delta S_{0}+\left(\frac{\partial \Delta S_{0}}{\partial T}\right)_{P}(T-T_{0})+\left(\frac{\partial \Delta S_{0}}{\partial P}\right)_{T}(P-P_{0}),
\end{equation} 
\begin{equation}\label{deltaV}
 \Delta V=\Delta V_{0}+\left(\frac{\partial \Delta V_{0}}{\partial T}\right)_{P}(T-T_{0})+\left(\frac{\partial \Delta V_{0}}{\partial P}\right)_{T}(P-P_{0}),
\end{equation} 
and 
from Eq.~(\ref{dG})--(\ref{deltaV}), by integration,
\begin{equation}
\begin{array}{cc}
\Delta G(P,T)=\frac{\Delta\beta}{2}(P-P_{0})^{2} + 2\Delta\alpha(P-P_{0})(T-T_{0}) + \\ \\ -\Delta C_{P}\left[(T-T_{0})-T_{0}\ln(T/T_{0})\right]+ \Delta V_{0}(P-P_{0}) -\Delta S_{0}(T-T_{0})+ \Delta G_{0},
\end{array}
\end{equation} 
where 
$\alpha=(\partial V/\partial T)_{P}=-(\partial S/\partial P)_{T}$
is the thermal expansivity
factor, related to the isobaric thermal expansion coefficient $\alpha_{P}$ by 
$\alpha_{P}=\alpha/V$;  
$C_{P}=T(\partial S/\partial T)_{P}$ is the isobaric heat capacity and 
$\beta=(\partial V/\partial P)_{T}$ is the isothermal compressibility
factor related to the isothermal compressibility $K_{T}$ by the
relation $K_{T}=-(\beta /V)$. All the quantities with the subscript
equal to zero are usually referred to ambient conditions. 
By developing the
logarithm to the second order around
$(T_{0},P_{0})$ 
\begin{equation}
 \ln\left(\frac{T}{T_{0}}\right)\sim \frac{T-T_{0}}{T_{0}}-\frac{\left( T-T_{0}\right)^{2}}{2T_{0}^{2}},
\end{equation}
we get 
\begin{equation}\label{taylor}
\begin{array}{cc}
\Delta G(P,T)=\frac{\Delta\beta}{2}(P-P_{0})^{2} + 2\Delta\alpha(P-P_{0})(T-T_{0}) + \\ \\ -\frac{\Delta C_{P}}{2T_{0}}\left(T-T_{0}\right)^{2}+ \Delta V_{0}(P-P_{0}) -\Delta S_{0}(T-T_{0})+ \Delta G_{0}
\end{array}
\end{equation}
that is the equation of en ellipse 
given the 
constraint 
\begin{equation}
 \Delta \alpha^{2}>\Delta C_{P}\Delta\beta/T_{0}  .
\end{equation} 
This condition is guaranteed by the different sign of $\Delta C_{P}$
and $\Delta \beta$, as can be observed 
 for
some proteins, as reported by Hawley \cite{hawley}.  

The Eq.~(\ref{taylor}) is $\Delta G(P,T)$ Taylor expansion arrested to 
the second order, holding for 
$\Delta \alpha$, $\Delta\beta$ and $\Delta C_{P}$ independent of $T$
and $P$, as generally valid.
Adding third order terms in the expansion makes minimal effects on the 
elliptic shape of the stability region.

At maximum pressure $P_{\rm max}$ of stability for the protein,
$d\Delta G/dT=\Delta S= 0$,
while at the maximum temperature $T_{\rm  max}$ of stability, $d\Delta G/dP=\Delta V= 0$.
Therefore, based on Hawley's theory it is possible to make general
predictions about the changes of $\Delta V$ and $\Delta S$ as
schematically summarized in Fig.~\ref{protein-phase-diagram}.
This phenomenological  theory has no explicit 
information about the protein structure, and makes
strong assumptions, such as, for
example, that the protein only has  
two states, or that equilibrium
thermodynamics holds during the denaturation. 
The last assumption, in particular, 
 implies that the all process would be reversible. 
Nevertheless, consistency with Hawley's theory is a good test for
models of protein unfolding. 
In the next section we review some of these models. The review does
not pretend to be exhaustive, but it has the aim of mentioning a
number of positive results of the theory of protein folding.

\section{Models for protein unfolding}

In 1989 Lau and Dill proposed the HP model for protein folding
\cite{lau}. 
By assuming that the exposed surface of hydrophobic residues is
energetically unfavorable at low $T$, the model reproduces the folding
of the protein (hydrophobic collapse). The protein is represented as a
self-avoiding chain on a lattice. The chain is composed by two
different categories of amino acids: H (non-polar) and P (polar). 
The presence of the aqueous environment is taken into account in
an effective way, by introducing an attractive contact interaction between H
monomers. No other interactions are present in the system. 

Under these hypothesis, the authors
show that the features of the folding process  
depend on the HH energy interaction, the length of
the chain, and the specific sequence of H and P monomers. Moreover,
for long chains one folded state dominates.

The model has the virtue to reduce the complexity of the folding
process  to a manageable level. All the electrostatic and
chemical properties of each amino-acid are simplified by allowing only
two possible states. 
The degrees of freedom of the solvent are not explicitly included and 
the driving force for the
folding is the hydrophobic interaction of non polar
monomers. Nevertheless, the HP model cannot describe cold denaturation.
Therefore, 
the experimental evidence of cold denatured proteins calls for a
reconsideration of the hydrophobic interaction and its dependence on
temperature and structure of hydration water \cite{nash, nash2,
  meersman2, goossens}. 

Back in 1948, Frank and Evans \cite{frank} discussed the tendency of
water to form ordered structures around non polar solutes to minimize
the free energy cost of solvation. As a consequence, hydrophobic solutes
are ``structure makers'' for water, facilitating the formation of
cages around the solute. The effect of these structures around
hydrophobic solutes is to reduce the entropy with respect to the bulk
 and to compensate, approximately, the enthalpy cost for the
creation of a cavity to allocate the solute.  

As discussed by Muller in 1990 \cite{muller}, the compensation of the
enthalpy implies that water-water hydrogen bonds (HBs) at the interface
with the hydrophobic solute are stronger than those in the bulk. This
 is consistent with the experimental observation that the excess molar heat 
capacity for a nonpolar  solute at infinite dilution in water  is
positive. This quantity, defined as 
the difference of the partial molar heat capacity in
solution with the heat capacity of the pure liquid solute, is
far larger at 25 $^{\circ}$C when the solvent is water
than for any other solvent \cite{muller, mirejovsky}.  

The statement that HBs are stronger at the hydrophobic interface has
led to the misconception that water around a hydrophobic solute has an
iceberg-like structure. Computer simulations \cite{geiger,
  vanbelle}, theoretical analysis \cite{lee2, lee3, madan}, and
neutral scattering studies \cite{finney} are inconsistent with
iceberg-like structures. Hence, the restructuring of water around a
solvent seems not to play  a relevant role in the hydrophobic
effect. Nevertheless, Muller \cite{muller} showed that if hydration HBs
are enthalpically stronger but fewer than in bulk, a model with
two-states HBs can reproduce the sign reversal of the 
proton NMR chemical shift with $T$ and the heat capacity
change upon hydration.

On the other hand, a common opinion \cite{bennaim, souza} is that  the large
free--energy change associated to the hydrophobic effect is due
to the small size of the water molecules with respect to the solutes,
and that the free--energy change
associated to the network reorganization around hydrophobic particles
is negligible due to
compensation of enthalpy and entropy, although it may account for the
large heat capacity change upon hydration.  This observation
apparently ruled out Muller model, where the enthalpy-entropy compensation
upon hydration was not present.

Nevertheless, Lee and Graziano in 1996 \cite{lee}  showed that 
Muller model can be
slightly modified to recover also the enthalpy-entropy compensation
upon hydration. 
The Muller-Lee-Graziano model was further simplified by De los
Rios and Caldarelli in 2000 \cite{rios,caldarelli,rios2} 
in order to reduce the number of parameter. 
By further simplifying the the description of bulk water, 
they recovered hot and cold
denaturation for a protein represented as a hydrophobic homopolymer.
A development of this model has been recently used 
to study the effective interaction between chaotropic agents and
proteins \cite{salvi}.  

The model by De los Rios and Caldarelli has been generalized by
Bruscolini and Casetti \cite{bruscolini, bruscolini2} in 2000 by
allowing each monomer of non-polar homopolymer to be in contact with a
cluster of water molecules. Each cluster has an infinite number of
possible states and only one state minimizes the free--energy cost of the
interaction with the hydrophobic monomers. The model reproduces the
trends of thermodynamics averages in accordance with experiments
\cite{makhatadze} and simulations \cite{silverstain}, and predicts the
warm and cold denaturation. These results are qualitatively similar to
those of the Muller-Lee-Graziano model, further supporting the
relevant role of the solvent in the folding-unfolding process.

Cold denaturation and  $T$-dependence of the hydrophobic effect were
also observed by
Dias et al. analyzing a nonpolar homopolymer in Mercedes-Benz (MB)
water \cite{dias}. The MB model, originally introduced by Ben-Naim
\cite{bennaim2}, represents water molecules as disks in two-dimensions with
three possible HBs (arms) as in a Mercedes-Benz logo. Water molecules
interact via van der Walls potential and HB interactions. HB
interaction is modeled with a Gaussian potential, favoring a fixed
value for the water-water distance and aligned arms for facing
molecules. Simulations show that the 
average HB energy is higher for shell water than for bulk water at
high $T$, while is lower at lower $T$. Therefore, by cooling the
solution, is energetically more convenient to increase the protein
surface exposed to water, inducing protein unfolding. 
In this model, the water molecules forming a cage around the protein monomers 
are strongly H-bonded to each
other. The highly ordered structure of the solvent around the monomers
decreases the entropy of water, compensating the increase of
the entropy associated to the protein unfolding. 

This model has been criticized
\cite{yoshidome} because it assumes, without proof, that the enthalpy
gain dominates at low $T$, giving rise to free--energy gain upon
unfolding of the protein. 
In particular, Yoshidome and Kinoshita \cite{yoshidome} analyzed by 
 integral equation theory  the behavior of a
nonpolar homopolymer  composed by fused hard-spheres of
different diameters immersed in smaller hard spheres, with permanent
electrostatic multiple moments, representing the solvent
\cite{Kusalik}. 
The protein--water
interaction is represented by a hard sphere potential and water--water
interaction by a 
hard sphere  potential and an electrostatic contribution given by the
electrostatic multipole expansion. The author found that
denaturation is characterized by large entropy loss and large enthalpy
gain. However, these two contributions to the free energy almost
completely cancel out and make no significant contribution to the
free-energy change. They found that the driving mechanism for cold 
denaturation is the translational entropic-loss of water due to the
large excluded volume of the hydrophobic particles. They observed that
at low $T$  water diffuses less,
therefore  the hydrophobic effect is weaker and the protein unfolds.  

\subsection{Pressure effects}

Pressure effects on protein denaturation have been also considered in 
microscopic theoretical models. For example, in 2003 Marqu\'es and coworkers
\cite{marques} considered a hydrophobic homopolymer, represented as
self-avoiding random walk, embedded in a water bath on a compressible
lattice model in two-dimensions. Water--water interactions are
represented by the Sastry et al. water model
\cite{sastry}. The polymer--water interaction is repulsive, being
proportional to the density number of HBs and to the number of the
missed native contact points of the protein. The model displays warm
denaturation, pressure denaturation and cold denaturation at high
pressure in agreement with stability diagram of some proteins
\cite{zhang}, although not with others \cite{lassalle}. A peculiarity of
this model is that the effective repulsion between protein and solvent
is coupled to the average number of HBs of bulk water. 

To remove this coupling to an average property of the bulk,
in 2007 Patel and coworkers \cite{patel} proposed a model
where water at
the interface with protein has a restricted number of accessible
orientations for the HBs compared to the bulk. Along with the entropic
cost, the interfacial HBs also have an additional enthalpic bonus with
respect to bulk water, following the ideas
discussed by Muller, Lee and Graziano. The model displays a stability
phase diagram with hot, cold and pressure denaturation. However, it
does not reproduce all the expected features the schematic phase
diagram of Fig.~(\ref{protein-phase-diagram}). In particular, the
model does not reproduce the elliptic shape of the phase diagram and
the low-$P$ region with $\Delta S>0$ and $\Delta V>0$ for hot
denaturation. These results were confirmed by extending the model to
the case with heteropolymers. 

In the attempt to reproduce the elliptic phase diagram
for protein stability, we propose here a model starting from the
assumption that HBs at the interface are
stronger compared to HBs in bulk water \cite{franzese6}.
We'll proceed as follows: in section 3 we describe the
model for nano-confined water, in section 4 we summarize recent 
results for the model, to clarify its water-like behavior. In
the section 5 we propose a protein-water interaction mechanism
displaying some preliminary results. 

\section{Hydrophobic nanoconfinement for water}

We consider a monolayer of water nano-confined between hydrophobic
plates. The interaction between water molecules and the surface is
represented by a hard-core repulsion. The confinement is such to
inhibit the formation of bulk water structures. For
example, bulk water is known to preferentially form four HBs with four
nearest neighbor molecules in an approximate tetrahedral structure at low
temperature and pressure \cite{soper}. Hard confinement inhibits the
formation of such bulk structure. For example, 
Kumar et al. \cite{kumar} found, by molecular dynamics simulations with
TIP5P-water confined between flat hydrophobic plates separated by
0.7~nm, an almost-flat monolayer of water molecules, each with four 
neighbors in an orientationally-disordered square lattice.

To define a tractable model we coarse-grain this structure of water under similar conditions
\cite{franzese,franzese2,franzese3,franzese4}. We
divide the volume between the hydrophobic plates, and 
accessible to water, into $N$ cells each containing one water molecule
and each with volume
$V/N=r^{2}h$, where $h$ is the separation between the flat planes,  
 $r>2r_{0}$ is the average distance between water molecules, with 
$r_{0}$ equal to the van der Waals radius of a water
molecule. 
The van der Waals attraction (due to dispersion forces) and
the repulsive interactions (due to the Pauli exclusion principle)
between water molecules are described by a Lennard-Jones interaction 
\begin{equation}
 U=-\sum_{ij}\epsilon\left[\left(\frac{r_{0}}{r_{ij}}\right)^{12}-\left(\frac{r_{0}}{r_{ij}}\right)^{6}\right]  ,
\end{equation} 
where $r_{ij}$ is the distance between molecules $i$ and $j$ and the sum
is performed over all the molecules.

To each cell we associate a variable $n_{i}=0,1$. If  the number
density $\rho_{i}$ of the cell $i$ is $\rho_{i}r^{2}h\geq0.5$ then
$n_{i}=1$, otherwise $n_{i}=0$. 

To take into account the decrease of
orientational entropy due to the formation of HBs, we introduce for
each water molecule $i$ four bonding indices  $\sigma_{ij}$, one for
each possible HB. Each variable $\sigma_{ij}$ can assume $q$ different
values, $\sigma_{ij}=1 ... q$. We choose the parameter $q$ by
selecting 30$^\circ$ as the maximum deviation from linear bond
(i.e. $q=180^\circ/30^\circ=6$). Hence, every molecule has $q^{4}=1296$
possible orientations. 

The covalent (directional) HB attraction component is
expressed by the Hamiltonian term 
\begin{equation}
\mathscr{H}_{\rm HB}=-J\sum_{<ij>}n_{i}n_{j}\delta_{\sigma_{ij}}\delta_{\sigma_{ji}}	 ,
\end{equation} 
where $J>0$ represent the covalent energy gained per HB, the sum is
over nearest neighbors cells, and
$\delta_{ab}=1$ if $a=b$, 0 otherwise. 

The experiments show that the formation of a HB leads to an open
structure that induces an increase of volume per molecule \cite{soper,
  debenedetti}. This effect is incorporated in the model by
considering that the total volume of the system is 
 \begin{equation}
 V\equiv V_{0}+N_{\rm HB}v_{\rm HB}  ,
\end{equation} 
where $V_{0}$ is the volume of the system without HBs, and
$v_{\rm{HB}}$ is the increment due to the HB.  

The term $\mathscr{H}_{\rm{HB}}$ quantifies the two-body component
for HB interaction. On the other hand, the distribution of O-O-O angle
shows a strong $T$-dependence \cite{ricci} that suggest the presence
of many-body component for HB interaction. We quantify this component by the
Hamiltonian term  
\begin{equation}
 \mathscr{H}_{\rm{Coop}}=-J_{\sigma}\sum_{i}\sum_{(k,l)_{i}}\delta_{\sigma_{ik}\sigma_{il}}	,
\end{equation}  
where $J_{\sigma}>0$ is the characteristic energy of the cooperative
component. The sum is performed over the six different pairs
$(k,l)_{i}$ of arms of the molecule $i$. Therefore the total Enthalpy
for the water is 
\begin{equation}\label{eq1}
 H=U+\mathscr{H}_{\rm{HB}}+\mathscr{H}_{\rm{Coop}}
 +PV=U-(J-Pv_{\rm{HB}})N_{\rm{HB}}-JN_{\sigma}	 ,
\end{equation} 
where 
\begin{equation}
  N_{\rm{HB}}\equiv\sum_{<ij>}n_{i}n_{j}\delta_{\sigma_{ij}}\delta_{\sigma_{ji}}	
\end{equation} 
is the total number of HB 
and
\begin{equation}
N_{\sigma}=\sum_{i}\sum_{(k,l)_{i}}\delta_{\sigma_{ik}\sigma_{il}}
\end{equation} 
is the total number of HBs optimizing the cooperative interaction.

\section{Results for water model}

We study our model by Monte Carlo (MC) simulations and mean-field
calculations. MC simulations are performed in the $NPT$ ensemble where
the volume of the system is a stochastic variable. We consider
periodic boundary conditions in the directions parallel to the
confining surfaces. 

\subsection{Liquid-gas phase transition and anomalies.}

Previous calculations have shown that the system displays a
liquid--gas first--order phase transition ending in a critical point $C$
at approximately $k_{B}T_{C}/\epsilon=1.9\pm0.1$ and
$P_{C}v_{0}/\epsilon=0.80\pm0.05$ \cite{franzese, franzese2,
  franzese3, franzese5, mazza, stokely}, in qualitative agreement with
mean field results \cite{franzese3, franzese4, stokely}. 

The model
reproduces several anomalies of water. For example, the system presents
density anomaly, i.e. the isobaric increase of density upon cooling,
up to reach a temperature of maximum density (TMD). The system also
displays diffusion anomalies \cite{dlsf},
maxima of isothermal compressibility $K_{T}$, isobaric heat
capacity $C_{P}$ and the isobaric thermal expansion coefficient
$\alpha_P$ \cite{franzese3, 
  franzese4, kumar2, kumar3, franzese9, mazza} related to the
anomalous behavior of water in the supercooled region. 
 
\subsection{Dynamical slowing down of water in supercooled region.}

The dynamical behavior of the model at low $T$ has interesting
features \cite{franzese5, kumar2, kumar3}. The dynamics of the HBs at
constant $P$ displays an increase of the correlation time when $T$ is
decreased. The increase is faster at higher $T$ then at lower $T$ and
shows a crossover at the temperature when $C_{P}$ has a maximum
\cite{kumar2, kumar3}. Results clarify that the crossover is due to a
structural change in the HBs network \cite{kumar2, kumar3}. The
qualitative features of this crossover have been successfully compared
to experimental results for confined water at increasing $P$
\cite{chu}. In particular Franzese and de los Santos \cite{franzese5}
have been showed that at high pressure ($P\simeq 2000$~bar) the
effect of HB is negligible due to 
the high enthalpic cost to form a HB and the correlation function
decays as an exponential. At low $P$ ($P\simeq 1$~bar) the correlation is
large also at long times and the system get stuck in a glassy
state. The structural analysis shows that under these conditions the
HB network develops gradually by decreasing $T$ and traps the system
in metastable configurations. For intermediate value of pressure the
correlation function $C(t)$ is well described by a stretched
exponential function 
\begin{equation}
 C(t)=C_{0}e^{-\left(\frac{t}{\tau}\right)^{\beta}} ,
\end{equation} 
where $C_{0}$, $\tau$ and $\beta\leq1$ are fitting constant ($\beta=1$
correspond to exponential decay). The quantity $1-\beta$ is a measure
of the heterogeneity in the system. As we approach to a characteristic
value of pressure $P_{C^{'}}$, $\beta$ reaches its minimum value
($\beta\simeq 0.4$). This is consistent with the experimental value of
$\beta\simeq 0.35$, observed for intermediate scattering
correlation function of water hydrating myoglobin at low hydration
level ($h=0.35$~g H$_{2}$O/g~protein) \cite{settles,
  doster}. Therefore, Franzese and de los Santos result indicates that
the system exhibits a largest amount of heterogeneity at
$P_{C'}$. As we will discuss in the next section, this heterogeneity
is the consequence of a large increase of cooperativity in the
vicinity of $P_{C'}$. 

\subsection{Thermodynamics of supercooled water.}\label{thermodynamics}

Four scenarios have been proposed to explain the
thermodynamics supercooled water. The \textit{stability limit}
scenario \cite{speedy} hypothesizes that the limit of stability of
superheated liquid water merges with the limit of stretched and
supercooled water, giving rise to a single locus in the $P-T$ plane,
with positive slope at high $T$ and negative slope at low $T$. The
reentrant behavior of this locus would be consistent with the
anomalies of water observed at higher $T$. As discussed by
Debenedetti, thermodynamic inconsistency challenges this scenario
\cite{debenedetti2}.  

The \textit{liquid-liquid critical point} (LLCP) scenario \cite{poole}
supposes a first order phase transition in the supercooled region
between two metastable liquids at different densities: the
low-density liquid (LDL) at low $P$ and $T$, 
and the high-density (HDL) at high $P$ and $T$.  
The phase transition line has  a negative slope in the $P-T$ plane and  
ends in a critical point. Numerical simulations
for several models are consistent with this scenario \cite{poole,
  poole2, tanaka, tanaka2}.  

The \textit{singularity-free} scenario \cite{sastry} focuses on the
anticorrelation between entropy and volume as cause of the large
increase of response functions at low $T$ and hypothesizes no HB
cooperativity. The scenario predicts lines of maxima 
in the $P-T$ for the response functions, similar to those observed in
the LLCP scenario, but shows no singularity for $T>0$. 

The \textit{critical-point-free} scenario \cite{angell} hypothesizes
an order-disorder transition, with a possible weak discontinuity of
density, that extends to $P<0$ and reaches the supercooled limit of
stability of liquid water. This scenario would effectively predicts no
critical point and a behavior for the limit of stability of liquid
water as in the stability limit scenario. 

As showed by Stokely et al. \cite{stokely}, all these scenarios may be
mapped into the space of parameter $J$ and $J_{\sigma}$, of the model
presented in the previous section, i.e. the coupling constants of the
covalent component of the HBs and the coupling constant of the many-body
component of the HB interaction, respectively. In particular, Stokely
et al. showed by mean field calculations and numerical simulations
that the absence of the many-body component leads to singularity free
scenario, while a large value of the many-body component with respect
to the covalent component give rise to the critical-point
free/stability limit scenario \cite{stokely}. 

By using estimate of 
these parameters from experimental results, the authors predict a
liquid-liquid phase transition at low temperature and high pressure
ending in a second critical point $C'$ for water
\cite{stokely}. Therefore, following this prediction, the increasing
fluctuations related to $C_{P}$, $K_{T}$ and $\alpha_{P}$ of water
under cooling are consequence of the liquid-liquid critical point
$C'$ in the supercooled region of water. By approaching $C'$,
the correlation length $\xi$ of the HBs increases. In particular for
any $P<P_{C'}$, the critical pressure, there is a temperature $T_{W}$
where the correlation length $\xi(P)$ is maximum. The locus
$T_{W}(P)$, called Widom line, converges toward $C'$ with a
negative slope in the $P-T$ plane \cite{franzese4, xu}. By increasing
$P$ along the Widom line, $\xi$ increases and diverges at
$P=P_{C'}$. Therefore, the regions of cooperativity of HBs increase in
size, leading to long cooperativity and, as a consequence, to larger
heterogeneity in the dynamics as observed
by Franzese and de los Santos \cite{franzese5} (discussed in the
previous section). 

Before discussing in more details the features
of these cooperative regions, is worth mentioning that recent
theoretical and experimental results on water hydrating lysozyme
proteins at very low hydration level
($h=0.3$~g H$_{2}$O/g protein) allow to explain the very low-$T$
phase diagram of water monolayer, at $T\simeq 150$~K at ambient
pressure \cite{mazza2}.  This investigation reveals that at low $P$
two structural changes take place in the HB network of the hydration
shell. One at about $250$~K is due to the building up of the HB
network \cite{kumar3}, and another at about $180$~K is
consequence of the cooperative reorganization of the HBs. These two
structural changes give rise to two dynamical crossovers in the HB
correlation time and the corresponding experimental quantity, the
proton relaxation time \cite{mazza2}. By increasing $P$, 
approaching $P_{C'}$, the two structural changes merge and at
$P_{C'}$ lead to diverging fluctuations associated to the liquid-liquid
critical point, as discussed in a recent work by Mazza et
al. \cite{mazza3}.

\section{Geometrical description of clusters of correlated HBs}

As discussed in the previous section, a water monolayer between
hydrophobic plates separated by less than $1$~nm, has a complex phase
behavior at low $T$ below the limit of stability of bulk liquid
water. The same phase diagram compares well with experiments with
water monolayer hydrating a complex substrate formed by proteins at
low hydration level \cite{settles, doster}. This can be understood if
we admit that the main effect of the protein substrate at low
hydration is to induce in the first layers of hydrating water a structure
that is inconsistent with any possible crystal. Therefore, the
substrate inhibits the crystallization, but does not inhibit the
water-water HB formation. It is, therefore, interesting to understand
how the region of correlated HB builds up and give rise to the
cooperative rearrangement and the liquid-liquid phase transition. 

To
this goal we follow an approach that has been validated during the
last three decades to describe critical phase transitions. It consists
in a percolation approach elaborated in 1980 by Coniglio and Klein
\cite{coniglio} for ferromagnetic systems and related to a mathematical
mapping introduced by Fortuin and Kasteleyn \cite{fortuin} and to
Swendsen-Wang \cite{swendsen} and Wolff \cite{wolff} techniques for
cluster MC methods. The Coniglio-Klein approach, called
{\em random-site-correlated-bond percolation}, was extended 
ferromagnetic systems with  many states \cite{coniglio2} and spin-glass-like systems
\cite{cataudella, franzese10, franzese11, franzese12}. In particular,
in Ref.~\cite{coniglio} it was proved that the clusters defined following
the rules of this specific type of percolation statistically coincide
with the region of thermodynamically correlated variables. Moreover,
in Ref.~\cite{franzese10} it was proved that this result holds as long
as the system has no frustration due to competing
interactions. Since in the case considered here there are no
competing interactions, we can follow the percolation approach to
define clusters of water molecules with statistically correlated HBs.  

As described in Ref.~\cite{franzese6}, we adopt the Wolff cluster MC
algorithm \cite{mazza} to study the cooperative regions and their
length scale. Thanks to the fact that a cluster represents a region of
water molecules with statistically correlated HBs, the algorithm allows
to equilibrate the system at any $T$ \cite{mazza}. 

By definition,
two variables $\sigma_{ij}$ and $\sigma_{ji}$ belong to the same
cluster with probability
\begin{equation}
p=\min\{\delta_{\sigma_{ij}\sigma_{ji}},1-\exp[-(J-Pv_{\rm{HB}})/kT]\}
\end{equation}
if they belong to nearest neighbor molecules, or with probability
\begin{equation}
p_{\sigma}=\min\{\delta_{\sigma_{il}\sigma_{ik}},1-\exp(-J_{\sigma}/kT)\}
\end{equation}
if they belong to the same water molecule $i$.  

The size of a cluster is given by the total number of $\sigma_{ij}$
variables belonging to the cluster. For each four $\sigma_{ij}$ 
in a cluster we have, on average, one water molecule in the
cluster. The average linear size of finite (non percolating) cluster
is, for the mapping discussed above, statistically equivalent to the
correlation length of the HBs. 
Moreover, it is possible to prove \cite{coniglio2} that  each
thermodynamic quantity, such as the compressibility, can be described
in terms of an appropriate moment of the finite cluster distribution. 

\begin{figure}
\centering
\includegraphics[scale=0.5]{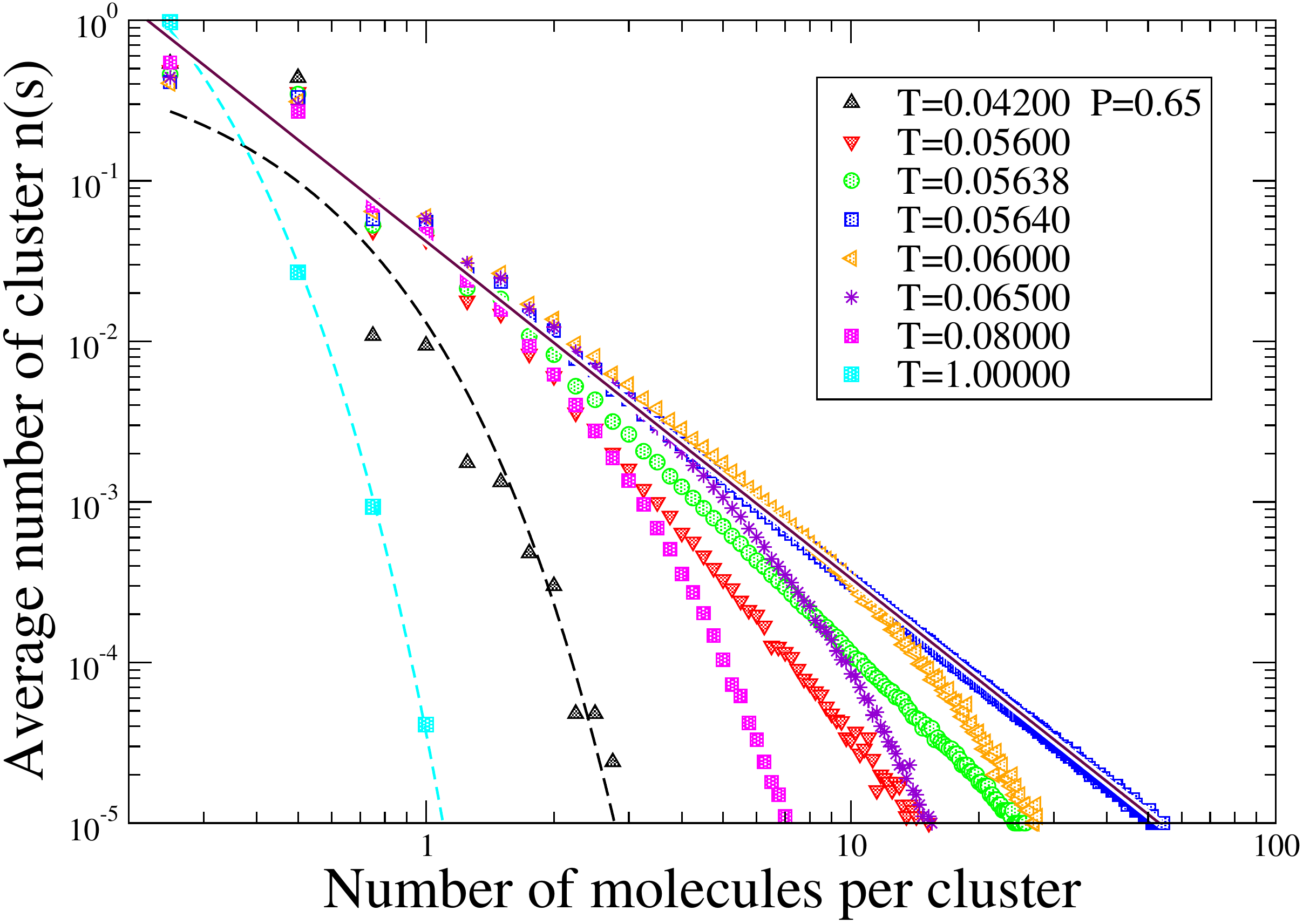}
\caption{Distribution $n(s)$ of clusters with finite size $s$
  formed by correlated
  hydrogen bonds (HBs) in a non-crystallizing water monolayer. Calculations are for  
$P=1.93$~GPa~$\simeq P_{C'}$, the liquid-liquid critical pressure, and
  different values of temperature for system with $N=4\times 10^4$ water
  molecules. 
For $T=173.84$~K~$\simeq T_{C'}$, the liquid-liquid critical
  temperature, calculations (blue square points) 
  decays as a power law $n(s)\sim s^{-\tau}$ with $\tau=2.1\pm 0.1$, as expected from
  theory near a critical point \cite{franzese10}.
Consistent with the theory, we find that 
$n(s)$ cannot be
described by a power law decay away from the critical point.
This is the case, for example, at $T=173.83$~K (green circles) and 
$T=176.35$~K (orange triangles).
We find that at temperature far from  the critical temperature,
$n(s)$ has an exponential decay, as expected.
This is the case, for example, at $T=833.08$~K  (light blue square)
and  $T=163.80$~K (black  triangles).
Continuous line is the power law fit, while dashed line are
exponential fits.
} 
\label{cluster-distribution}
\end{figure}

By approaching the critical point $C'$, we observe that the largest
cluster percolates and its linear size becomes comparable to the
system size. Under these conditions, the correlation length $\xi$
diverges. While away from $C'$ the distribution $n(s)$ of cluster of
linear size $s$ decays as an exponential,
$n(s)$ has
a power law decay near  $C'$ (Fig. \ref{cluster-distribution})
consistent with the theory \cite{franzese10}.

From general considerations it is possible to show that $n(s)\sim
s^{-\tau}$, where the exponent $\tau$ is related to the fractal
dimension $D_{F}$ of the system $\tau=1+d/D_{F}$, and  $d=2$ is the
embedding (euclidean) dimension. A preliminary estimate $\tau\simeq 2$
suggests that the clusters of correlated HBs are compact with $D_{F}\sim2$
\cite{bianco}. Therefore, the mapping of the thermodynamic systems into
a percolation problem allows us to give a geometrical description of the
correlated HBs.  

\section{Free energy landscape analysis}

The percolation approach allows us to adopt  a cluster MC dynamics that is
very efficient at low $T$ \cite{mazza}. Therefore, we can equilibrate
the system at low $P$ around the $T_{W}(P)$, the temperature  of maximum correlation
length $\xi$, and 
 around the temperature $T_{LL}$ of liquid-liquid
coexistence at high $P$, and calculate the free
energy landscape for the system. 

\begin{figure}
\centering
\includegraphics[scale=1]{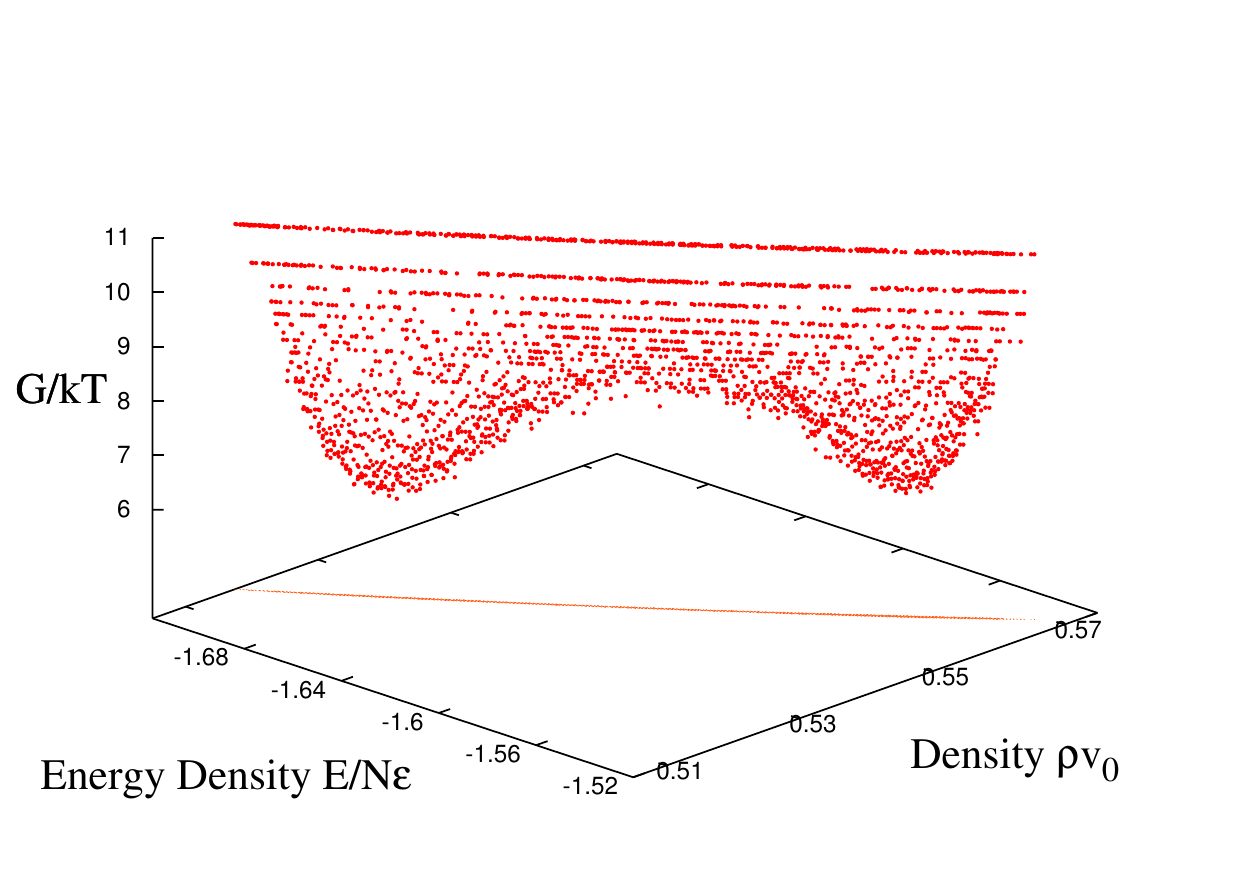}
\caption{Gibbs free energy for a water monolayer with $N=4\times 10^4$ water molecules
at $P=1.98$~GPa  and $T=158$~K. The two minima, one
  at low energy and density, and the other at high energy and density,
  represent respectively the LDL phase and HDL phase of the
  system. The projection of $G$ on the plane $E/N-\rho$ shows that
  there is a linear relation between the accessible energies and
  density for the coexisting states.
The same value of the minima marks 
the coexistence of the two phases.} 
\label{free-energy}
\end{figure}

By definition,
the Gibbs free energy is 
\begin{equation}
G/k_BT\equiv -\ln \mathscr{P}(H,\rho),
\end{equation}
 where $\mathscr{P}(H,\rho)$ is
the density of states with enthalpy $H\in[H,H+\delta H]$ and density
$\rho\in[\rho,\rho+\delta \rho]$, with $\delta H$ and $\delta\rho$ 
infinitesimal increments. In Fig.~\ref{free-energy} we show 
$G$ as a function
of energy per particle $E/N$ and the density
$\rho$ with two equivalent  minima 
straddling the line of phase
transition. The two minima, equivalent 
within the numerical precision, correspond to two coexisting phases with
different density and energy.
The one at higher
density and energy represents the HDL phase.
The other at lower density and energy represents the
LDL phase. Approaching $C'$ the two minima get closer and the density
separation disappears, as expected at the critical point. These
results are consistent with the mean field free energy analysis of
Ref.~\cite{mazza, franzese8}, where the Gibbs free energy is calculated
as 
function of the HB order parameters relevant at the structural
transition at high $P$ and low $T$. In the mean field analysis the
minima of $G$ are separated at high $P$, but merge for $P$ approaching
$P_{C'}$. All these results are consistent with the behavior of
$C_{P}$, $K_{T}$ and $\alpha_{P}$ whose maxima move to lower $T$ as
$P$ is increased \cite{kumar2, franzese3, kumar2, franzese9}. The loci
of the maxima of $C_{P}$, $K_{T}$ and $\alpha_{P}$ merge in the
vicinity of $C'$ and the
amplitude of their maxima increases approaching $C'$. 

\section{A model for protein in water}

In the previous section we define a coarse-grained model for a
water monolayer and we show that the model compares well with
experiments for protein shell--water and that it predicts a complex
phase diagram for low $T$, below the limit of stability of bulk liquid water,
and  high $P$. As discussed in the introduction, under these 
 conditions folded proteins are
destabilized. Following our 
discussion about how could be relevant to
take into account the HB free energy 
to explain the lost of stability of folded
proteins, It is intriguing to test if the proposed water
model could give insight into the mechanism of unfolding. 

To this goal
we modify the water model  to introduce the
effect of the protein-water interface. For sake of the  simplicity, 
we will limit our discussion to the case of a single protein
embedded into a water monolayer. Although far from the complex
studies of a protein embedded into bulk water, 
the model gives instructive results.  

\begin{figure}
\centering
\includegraphics[scale=0.2]{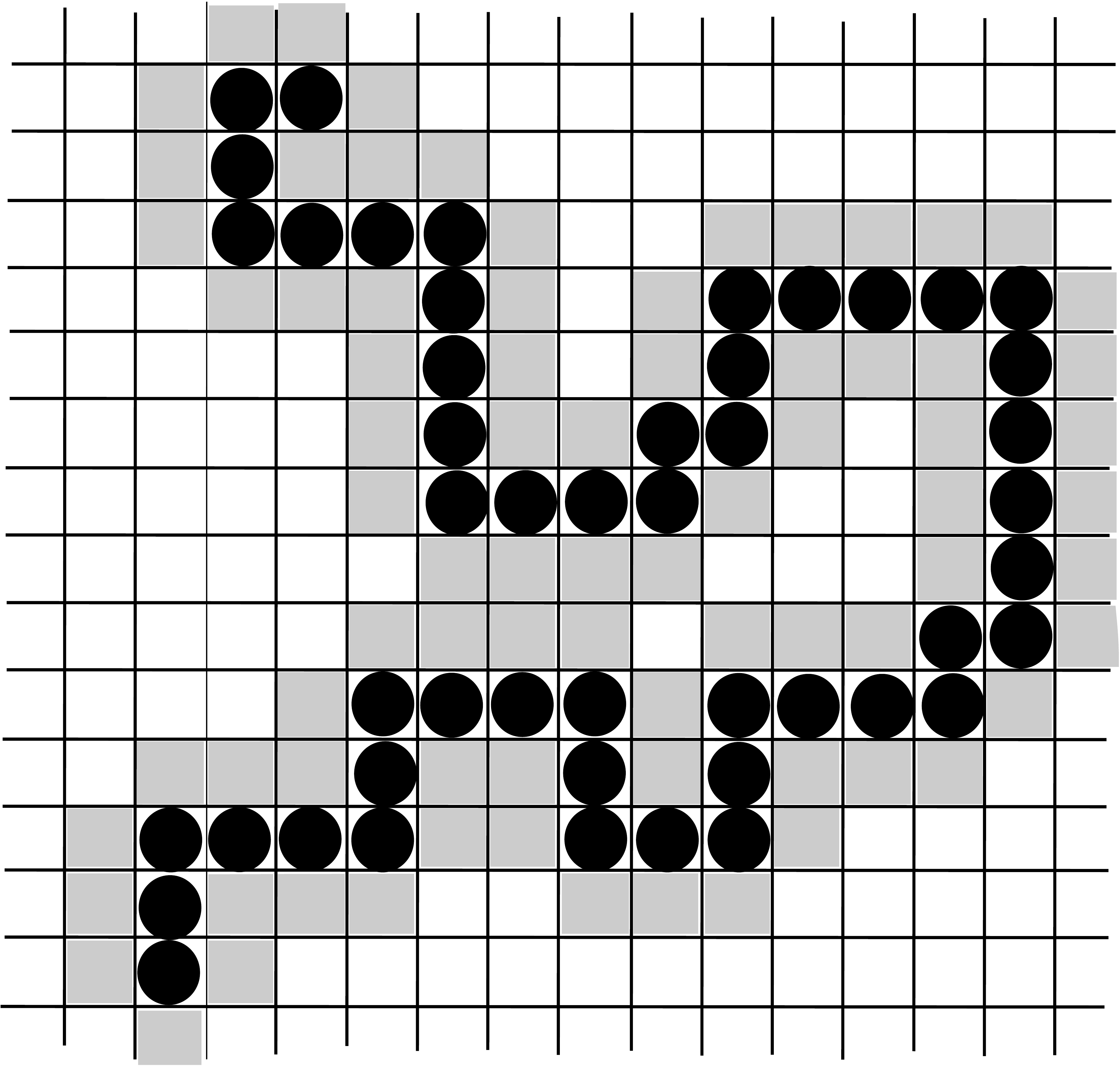}
\caption{Example of configuration of a homopolymer in the
  coarse-grained model. Each cell is occupied either by a water
  molecule (white and gray cells) or a hydrophobic homopolymer
monomer (cells with a full black circle). The gray cells represents
the sites occupied by shell water. The enthalpy gain for HB formation between shell
water molecules is larger than that between bulk water molecules,
according to the Eq.~\ref{hamiltonian-lambda}. Shell water molecules 
cannot form hydrogen bonds with nearest neighbors hydrophobic monomers.}  
\label{protein}
\end{figure}

The simplest protein that we can consider is a hydrophobic
homopolymer, schematized as a self avoiding chain
(Fig.~(\ref{protein}). Following the discussion by Muller
\cite{muller}, we 
require that, consistent with experiment, water molecules in contact
with a hydrophobic monomer have larger decrease of enthalpy upon HB
formation. Also consistent with Muller-Lee-Graziano discussion
\cite{lee}, the fraction of broken HBs at the hydrophobic interface is
larger than the fraction of broken HBs in the bulk. 

The first requirement is achieved by adding a term to the water Hamiltonian Eq.~(\ref{eq1})
\begin{equation}\label{hamiltonian-lambda}
 \mathscr{H}_{\rm s}=-\lambda J\sum_{<ij>_{\rm s}}n_{i}n_{j}\delta_{\sigma_{ij}}\delta_{\sigma_{ji}},
\end{equation} 
where the sum is taken over 
nearest neighbor water molecules in the protein hydration shell (Fig.~\ref{protein}),
and $\lambda>0$ is an adjustable parameter accounting for the larger
enthalpy decrease for HBs in the hydration shell. Hence, for HB
formed between water molecules in the shell, the enthalpy
variation is $-J(1+\lambda)+Pv_{\rm HB}$, and the total enthalpy
for protein into water is 
\begin{equation}
H_{\rm tot}=H+\mathscr{H}_{\rm s},
\end{equation} 
where $H$ is given by Eq.~(\ref{eq1}).

The second requirement of Muller-Lee-Graziano approach, i.e. a larger
number of broken HBs at the interface, is achieved by volume
exclusion. Once a cell of our system is occupied by a protein monomer,
it cannot be occupied by a water molecule. Therefore, a water
molecule, in the hydration shell cannot form the HB in the direction
of the monomer and looses at least one HB (it can loose more if it has
more monomers as nearest neighbors, as shown in
Fig.~\ref{protein}). 

In the following section we will define the
algorithm 
adopted to generate protein equilibrium configurations. To this
propose we follow a MC procedure that mimics the dynamics at large
time scales.  

\subsection{Monte Carlo algorithm}

\begin{figure}
\centering
\includegraphics[scale=0.15]{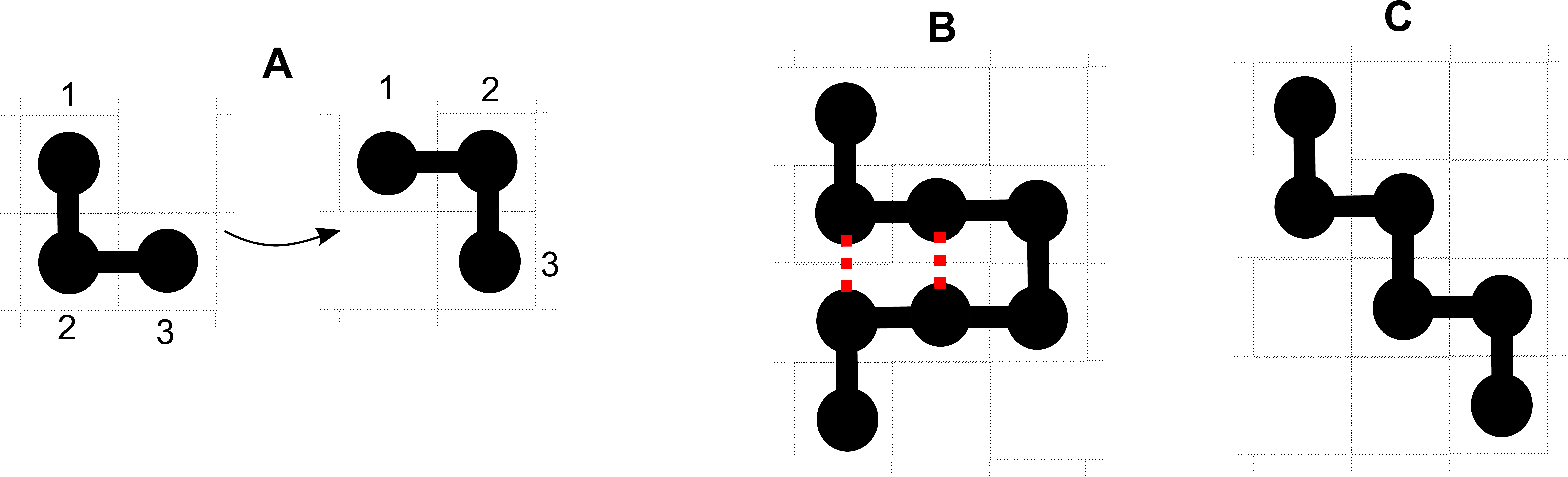}
\caption{A) Monomer 2 is in a corner configuration and can be
  displaced from the configuration on the left to the configuration on
  the right and {\it vice versa}.
B) Homopolymer configuration with two contact points, indicated with dotted lines. 
C) Homopolymer configuration without contact points.}
\label{displacement}
\end{figure}

We perform MC simulations in the $NPT$ ensemble. In every MC
step we choose a cell at random. If it is occupied by a water molecule,
we change randomly one of its $\sigma$ variables. If it is occupied by
a monomer and if the monomer is in a corner configuration
(Fig.~\ref{displacement}A) then we swap its position with the position of
the water 
molecules in the cell in the opposite corner. By doing this, we
keep the inter-monomer distances constant. 

If the cell, picked at random, is occupied by a monomer not in a corner
configuration, no displacement is performed because it would change
the inter-monomer distance. This limitation is introduced in order to
avoid in the free energy any term accounting for the
elastic energy of the homopolymer. The effects of this elastic
contribution are for the moment outside of the scope of the present
work.  

Finally, as in the cases discussed in the previous section, to keep
the pressure of the system constant, every $N$ random changes of the
cell variables (where $N$ is the total number of cells in the system),
we attempt to rescale all the system volume by a factor that is tuned
in a way to guaranty 50\% of acceptance ratio. All the MC moves described
above are accepted or rejected according to the Boltzmann factor
associated to the enthalpy change caused by the move. 

In order to study the folding-unfolding process of the proteins we
calculate the number of contact points $N_{\rm cpts}$ as
illustrated in Fig.~(\ref{displacement}). In this calculation we do
not count the monomers that are adjacent along the homopolymer.  

\section{Preliminary results for hydrated homopolymers}

We study a system with $N$ water molecules and a hydrophobic
homopolymer chain with $N_{\rm m}$ monomers. In our preliminary
simulations we used $N=650$ or $N=1000$ and $N_{\rm m}=12$ or 
$N_{\rm m}=50$. The parameters are chosen, for consistency, as in
previous analysis \cite{strekalova}: $\epsilon=5.8$~kJ/mol,
$J=2.9$~kJ/mol, $J_{\sigma}=0.29$~kJ/mol, $v_{0}=hr_{0}^{2}$, $h=7$~\AA,
$v_{\rm{HB}}/v_{0}=0.5$ and $q=6$. We choose $\lambda=0.7$ for the
larger decrease of enthalpy at the hydrophobic interface.

Our results display a non monotonic behavior of
$N_{\rm cpts}$ as function of $T$ at low $P$. At high pressure
we observe an region in the $P-T$ plane where the number of 
contact points is 
at least 51\% of the maximum possible number. By
definition, we consider these configurations as belonging to the set
of folded states. We observe that the region of folded states is
included within a larger region in the $P-T$ plane where the number of 
contact points is 
at least 49\% of the maximum possible number. 
By definition, we consider those configurations as members of
the set of state representing the molten globule \cite{Vendruscolo}.

\begin{figure}
\centering
(a) \hspace{2.8cm}
(b) \hspace{2.8cm}
(c) \hspace{2.8cm}
(d) \hspace{2.8cm}
(e) \hspace{2.8cm}
\includegraphics[scale=0.15]{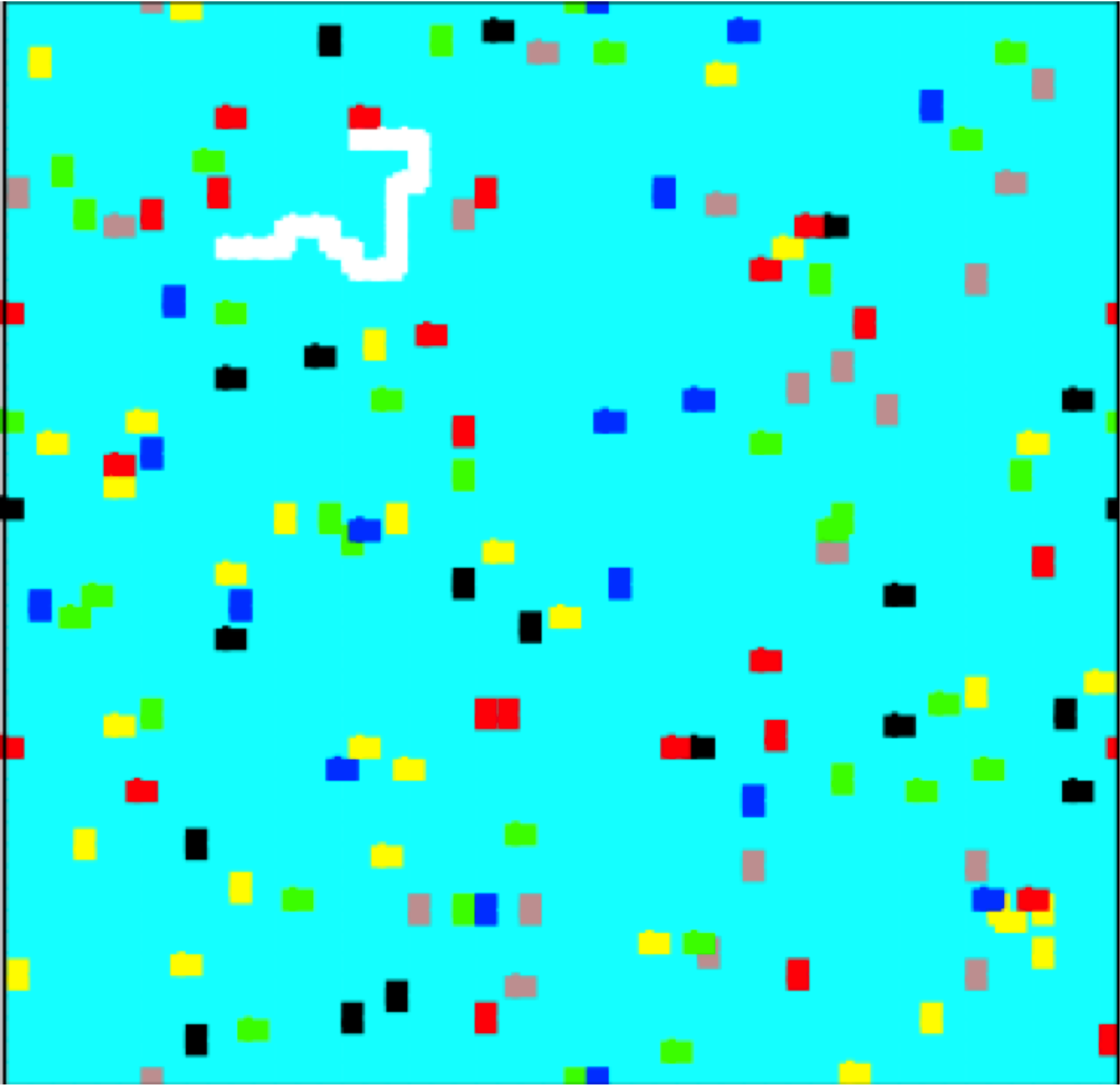}
\includegraphics[scale=0.15]{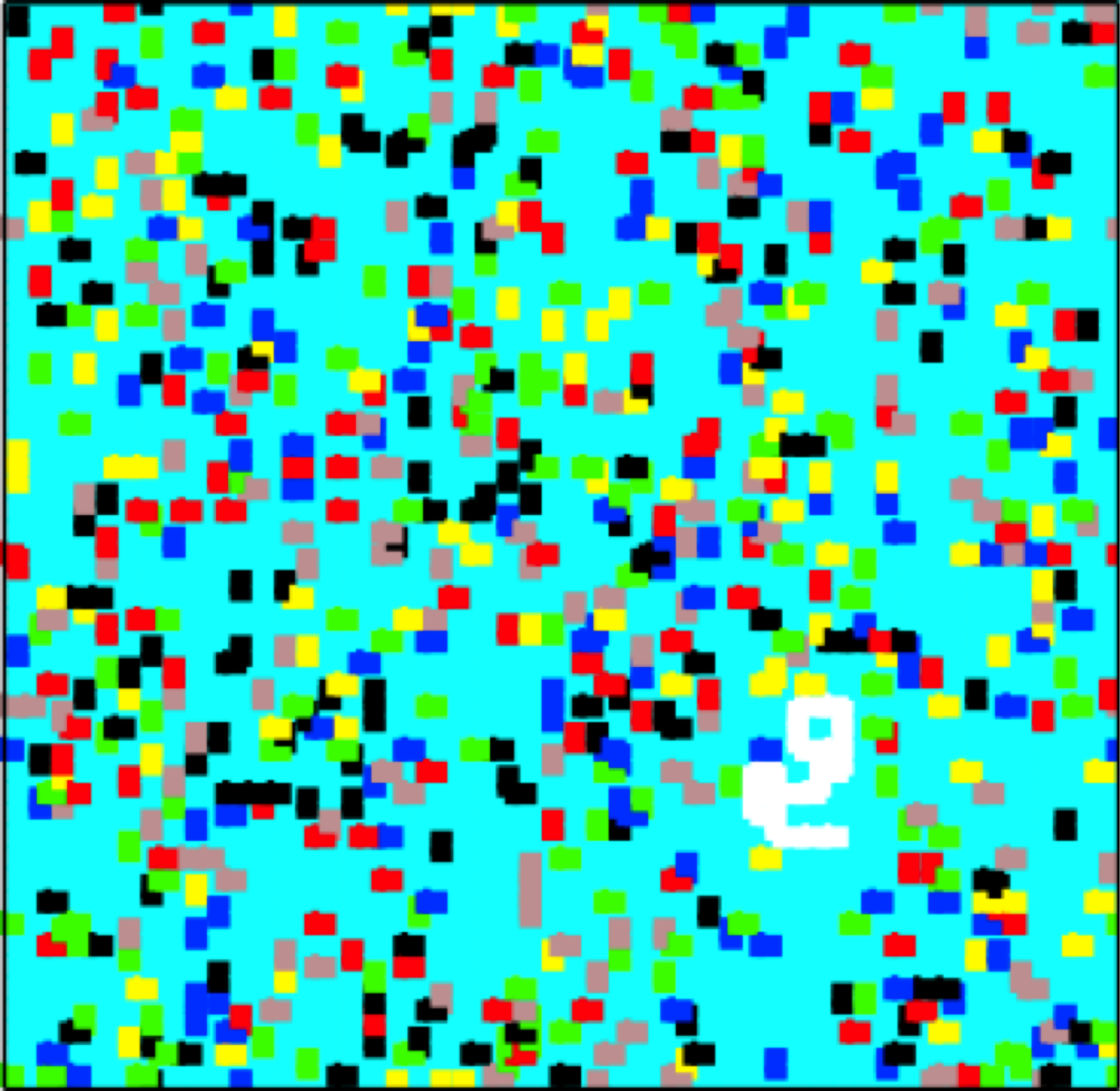}
\includegraphics[scale=0.15]{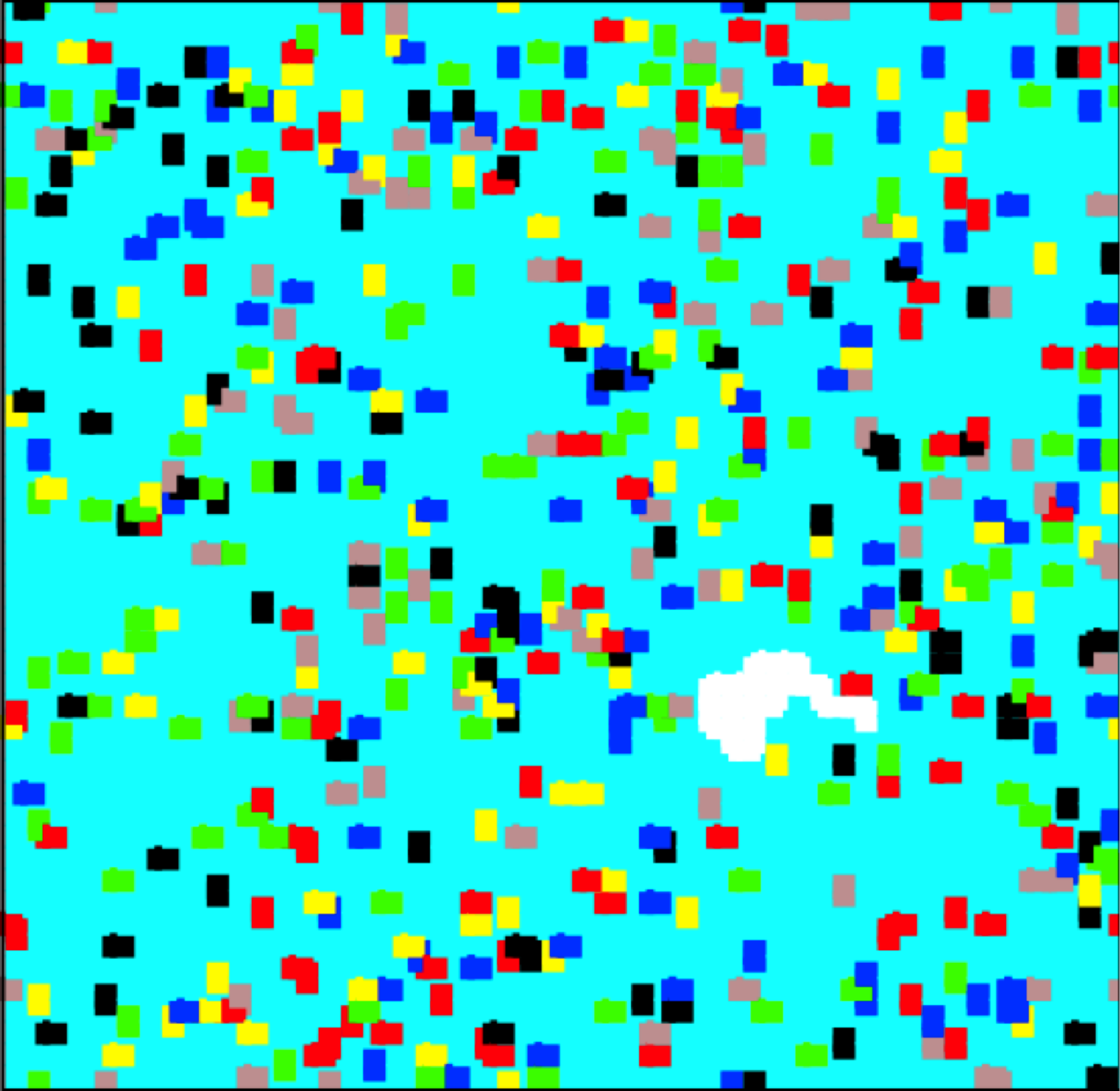}
\includegraphics[scale=0.15]{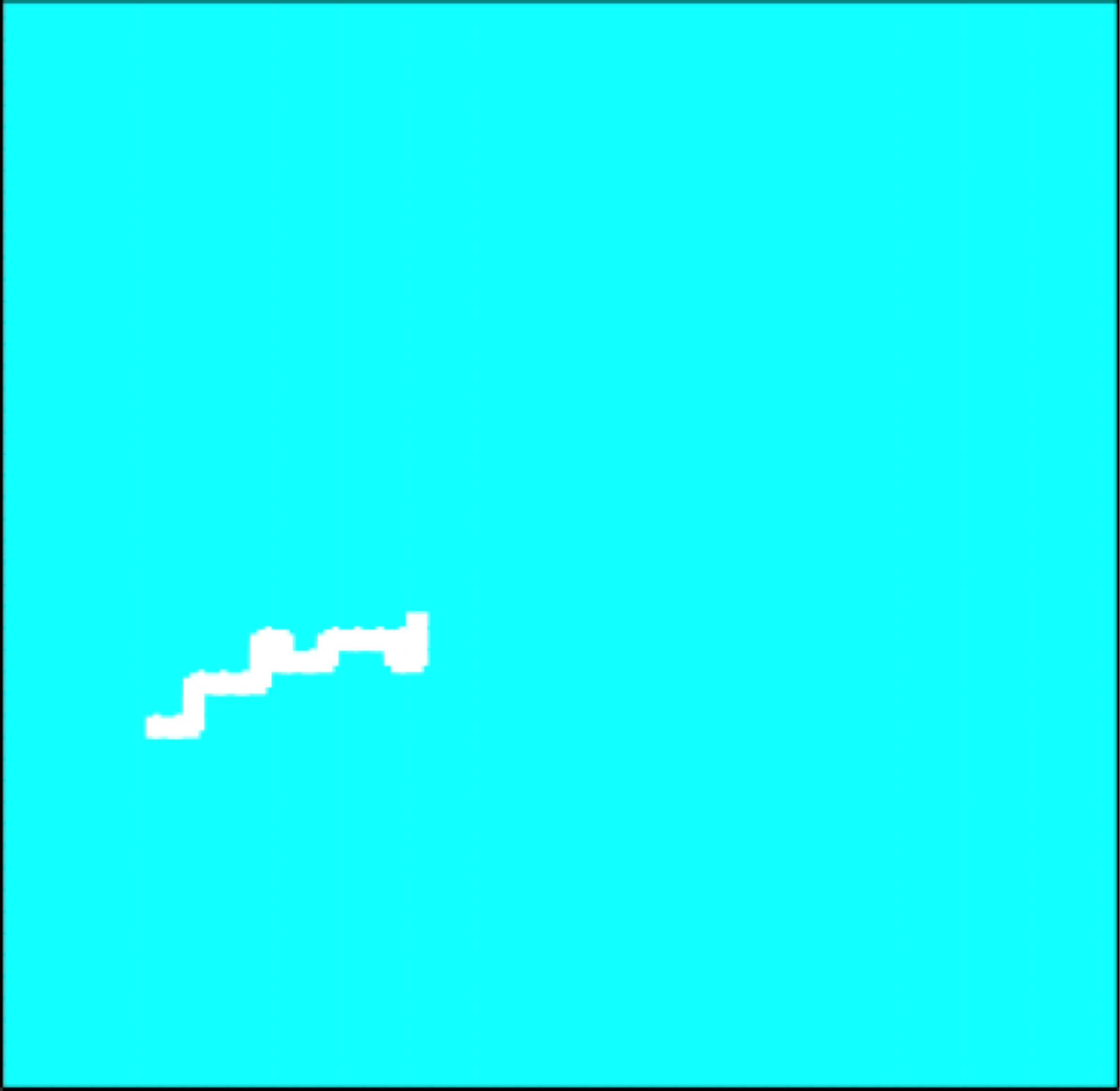}
\includegraphics[scale=0.15]{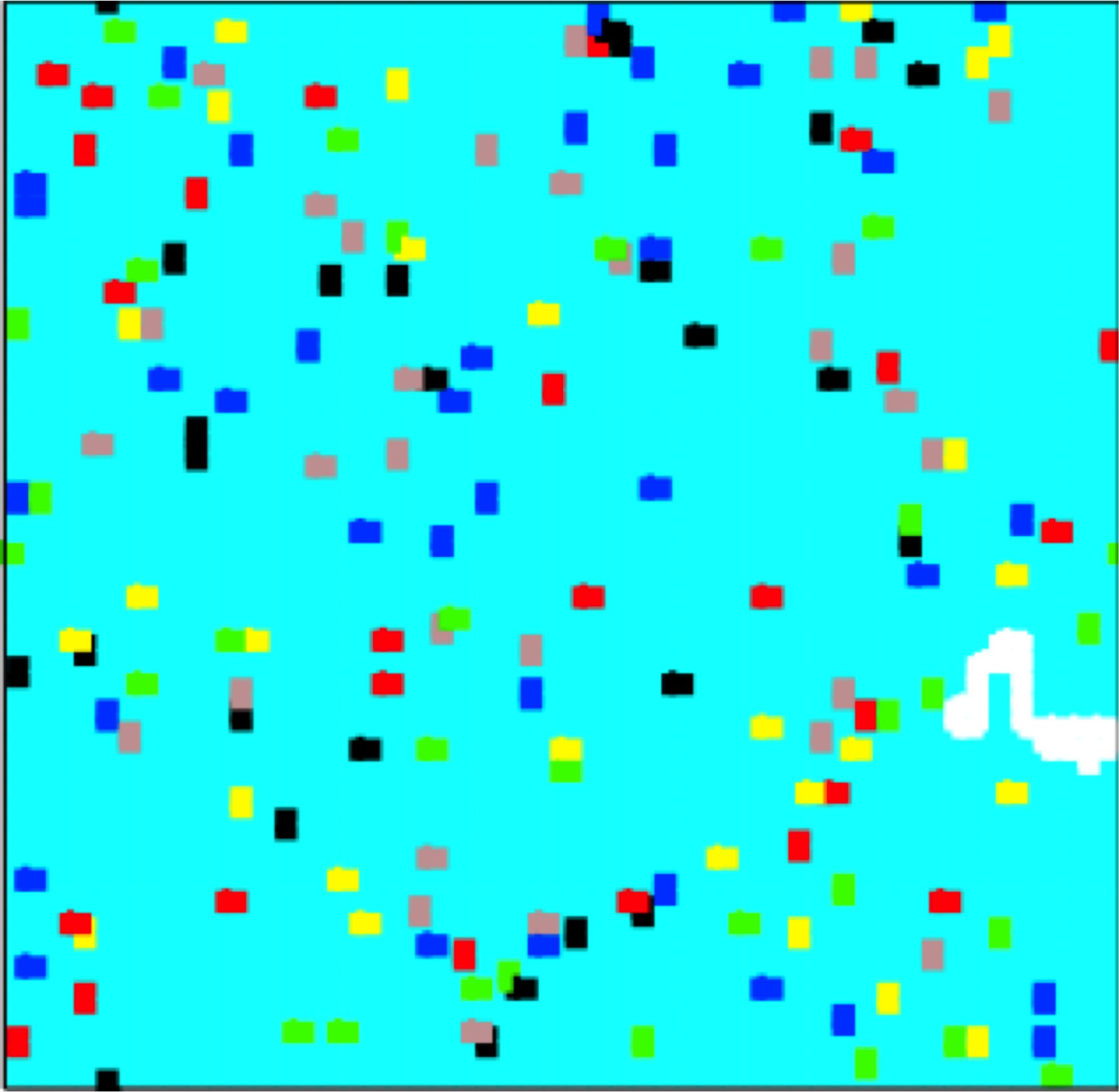}
\caption{Typical configurations of
folding--unfolding of a coarse-grained protein suspended
  in water at different temperatures $T$ and high pressures $P$. 
The protein is
  represented as a fully hydrophobic chain (in white), surrounded by
  water molecules (turquoise background). We use different color for
  water to indicate the different orientations of the HBs.  
(a) At high $P$ and high $T$, the protein unfolds and the number of
HBs (colored sticks) of surrounding water is small.  
(b) At the same pressure but lower $T$, the protein collapses in a
molten globule state. (c) At lower $T$ the protein folds, while the
surrounded water has a large number of HBs. (d) At much lower $T$ we
observe cold denaturation of the protein when the number of HBs
is largely reduced (zero in the configuration represented here).  
(e) At higher $P$ the denaturation occurs at higher $T$,
and the mechanism of unfolding seems to be dominated by the reduction
of HBs also under these conditions.} 
\label{protein-folding}
\end{figure}

We find that the region of folded states has an
elliptic shape that resembles the theoretical prediction
(Fig.~\ref{protein-phase-diagram}). 
In
particular, we observe that a folded protein unfolds upon cooling,
giving rise to the cold denaturation process. It also unfold by
increasing the pressure as expected by pressure denaturation
(Fig.~\ref{protein-folding}). 
Since
our stability region is at high $P$, we are also able to observe the
unfolding by decreasing the pressure, a phenomena that is predicted by
general theoretical considerations, as discussed in the introduction.
We also find that the axes of the elliptical stability
region are tilted as expected (Fig.~\ref{protein-phase-diagram}).

\section{Conclusions}

The behavior of supercooled water is still under debate and the
presence of a second critical point $C'$ could be relevant to
understand how the structure of liquid water changes around
proteins and how affects protein properties. Experiments of water
confined in nano-structures 
offer the possibility to access a range of temperatures where 
bulk liquid water would not be stable and would form ice. 
Hence, confinement allows to study water under conditions that are 
relevant in biological systems. 

Despite the growing interest of the scientific community in water at
 hydrophobic and hydrophilic interfaces, it is still unclear how the
 interaction with the confining 
surfaces affects the thermodynamics of water. For example, recently
Strekalova et al. \cite{strekalova} observed that 
the fluctuations of supercooled water confined into a hydrophobic
porous material are drastically smaller than those of bulk water. They
found that the response functions $C_{P}$, 
$\alpha_{P}$ and $K_{T}$ are largely reduced as a consequence of the
interaction with the porous medium. An extreme consequence of this
change is the disappearing of the liquid-liquid phase transition at
high pressures \cite{strekalova}.  
Therefore, 
further work is necessary to clarify the many issues
related to the dynamics and thermodynamics of water at the
interfaces. 

Here we presented a coarse-grained model for a monolayer of water and
its extension to the case of solvated proteins. The model takes into
account the cooperativity between HBs and has been studied by
simulations and mean field calculations. Previous results about the 
phase diagram, the diffusivity properties, the response functions
$C_{P}$, $\alpha_{P}$ and $K_{T}$ of the model and the connection of
these quantities  with the HBs
dynamics are in agreement with experimental results and validate the
model.  

We adopted this model in the context of protein folding. For the sake
of simplicity we consider the case of a protein schematized
as a
self-avoiding  hydrophobic homopolymer. Following  Muller analysis
\cite{muller}, we assume that the network of HBs is perturbed by the
presence of hydrophobic solute with large size. 

Our preliminary results reproduce hot, cold and pressure denaturation as well
as the existence of intermediate states (molten globule). We find that
the stability region for folded protein has the theoretically expected
elliptic shape in the $P-T$ plane. Further work is in progress to
elucidate the relevant  mechanism ruling protein stability. 
\\\\

\section*{Acknowledgments}

We thank for enlightening discussion 
G. Caldarelli, 
P. de los Rios, 
P. G. Debenedetti, 
C. M. Dobson, 
M. Vendruscolo.
G.F. thanks for collaboration and helpful discussions
M. C. Barbosa,
S. V. Buldyrev,
F. Bruni,
S.-H. Chen,
A. Hernando-Mart\'{\i}nez,
P. Kumar, 
G. Malescio,
F. Mallamace,
M. I. Marqu\'es,
M. G. Mazza,
A. B. de Oliveira,
S. Pagnotta,
F. de los Santos,
H. E. Stanley, 
K. Stokely,
E. G. Strekalova,
P. Vilaseca,
and the Spanish Ministerio de Ciencia e Innovaci\'on
Grants FIS2009-10210 (co-financed FEDER)
for support.


\bigskip

\begin{thebibliography}{10}
\bibitem{ravindra} Ravindra R, Winter R, On the temperature-pressure free-energy landscape of proteins, Chem Phys Chem 2003;4:359-365.

\bibitem{pastore} Pastore A, Martin S R, Politou A, Kondapalli K C, Stemmeler T, Temussi P A, Unbiased Cold Denaturation: Low- and High-Temperature Unfolding of Yeast Frataxin under Physiological Conditions, J Am Chem Soc 2007;129:5374-5375.

\bibitem{privalov} Privalov P L, Cold Denaturation of Proteins, Crit Rev Biochem Mol Biol 1990;25:281-305.

\bibitem{nash} Nash D, Jonas J, Structure of the pressure-assisted cold denatured state of ubiquitum, Biochem Biophys Res Commun 1997;238:289-291.

\bibitem{nash2} Nash D, Jonas J, Structure of the pressure-assisted cold denatured lysozyme and comparison with lysozyme folding intermediates, Biochemistry 1997;36:14375-17383.

\bibitem{meersman2} Meersman F, Smeller L, Heremans K, Pressure-assisted cold unfolding of proteins and its effects on the conformational stability compared to pressure and heat unfolding, High Press Res 2000;19:263-268.

\bibitem{goossens} Goossens K, Smeller L, Frank J, Heremans K, Pressure tuning spettroscopy of bovin pancreatic trypsin inhibitor: a high pressure FT-IR study, Eur J Biochem 1996;236:254-262.

\bibitem{meersman} Meersman F, Dobson C M, Heremans K, Protein unfolding, amyloid fibril formation and configurational energy landscapes under high pressure conditions, Chem Soc Rev 2006;35:908-917.

\bibitem{hummer} Hummer G, Garde S, Garcia A E, Paulaitis M E, The pressure dependence of hydrophobic interaction is consistent with the observed pressure denaturation of proteins, Proc Natl Acad Sci USA 1998;95:1552-1555.

\bibitem{kunugi} Kunugi S, Yamamoto H, Makino M, Tada T, Uehara-Kunugi Y, Pressure-assisted cold-denaturion of carboxipeptidase Y, Bull Chem Soc Jpn 1999;72:2803-2806.

\bibitem{hawley} Hawley S A, Reversible pressure-temperature denaturation of chymotrypsinogen, Biochemistry 1971;10:2436-2442.

\bibitem{smeller} Smeller L, Pressure-temperature phase diagrams of biomolecules, Biochim Biophys Acta 2002;1595:11-29.

\bibitem{lau} Lau K F, Dill K A,  A lattice statistical mechanics model of the conformational and sequence spaces of proteins, Macromolecules 1989;22:3986-3997.

\bibitem{frank} Frank H S, Evans M W, Free Volume and Entropy in Condensed Systems III. Entropy in Binary Liquid Mixtures; Partial Molal Entropy in Dilute Solutions; Structure and Thermodynamics in Aqueous Electrolytes, J Chem Phys 1945;13:507-533.

\bibitem{muller} Muller N, Search for a realistic view of hydrophobic effects, Acc Chem Res 1990;23:23-28.

\bibitem{mirejovsky} Mirejovsky D, Arnett E M, Heat capacity of the solution for alcohols in polar solvents and the new view of hydrophobic effects, J Am Chem Soc 1983;105:1112-1117.

\bibitem{geiger} Geiger A, Rahman A, Stillinger F H, Molecular Dynamics Study of the Hydration of Lennard-Jones Solutes, J Chem Phys 1979;70:263-276.

\bibitem{vanbelle} van Belle D, Wodak S J, Molecular dynamics study of methane hydration and methane association in a ploarizable water phase, J Am Chem Soc 1993;115:647-652.

\bibitem{lee2} Lee B, In An anatomy of hydrophobicity; Eds: Eisenfeld J, DiLisi C; Elsevier, North-Holland: Amsterdam, 1985.

\bibitem{lee3} Lee B, Solvent reorganization contribution to the transfer thermodynamics of small nonpolar molecule, Biopolymers 1991;31:993-1008.

\bibitem{madan} Madan B, Lee B, Role of hydrogen bonds in hydrophobicity: the free energy of cavity formation in water models with and without the hydrogen bonds, Biophys Chem 1994;51:279-289.

\bibitem{finney} Finney J L, Soper A K, Solvent structure and perturbations in solutions of chemical and biological importance, Chem Soc Revs 1994;1:1-10.

\bibitem{bennaim} Ben-Naim A, Hydrophobic interaction and structural changes in the solvent, Biopolymers 1975;14:1337-1355.

\bibitem{souza} Souza L E S d, Ben-Amotz D J, Hard Fluid Model for Molecular Solvation Free Energies, J Chem Phys 1994;101:9858-9863.

\bibitem{lee} Lee B, Graziano G, A Two-State Model of Hydrophobic Hydration That Produces Compensating Enthalpy and Entropy Changes, J Am Chem Soc 1996;22:5163-5168.

\bibitem{rios} Rios P D L, Caldarelli G,  Putting proteins back into water, Phys Rew E 2000;62:8449-8452.

\bibitem{caldarelli} Caldarelli G, Rios P D L, Cold and Warm Denaturation of Proteins, J Biol Phys 2001;27:229-241.

\bibitem{rios2} Rios P D L, Caldarelli G, Cold and Warm Denaturation of Hydrophobic Polymers, cond-mat/9903394v3 2008.

\bibitem{salvi} Salvi G, Rios P D L, Vendruscolo M, Effective Interactions Between Chaotropic Agents and Proteins, Proteins: Structure, Function, Bioinformatics 2005;261:492–499.

\bibitem{bruscolini} Bruscolini P, Casetti L, Lattice model for cold and warm swelling of polymers in water, Phys Rev E 2000;61:2208-2211.

\bibitem{bruscolini2} Bruscolini P, Casetti L, Bethe approximation for a model of polymer solvation, Phys Rev E 2001;64:051805.

\bibitem{makhatadze} Makhatadze G I, Privalov P L, Energetics of protein structure, Adv Protein Chem 1995;47:307-425.

\bibitem{silverstain} Silverstain K A T, Haymet A D J, Dill K A, A Simple Model of Water and the Hydrophobic Effect, J Am Chem Soc 1998;120:3166-3175. 

\bibitem{dias} Dias C L, Ala-Nissila T, Karttunen M, Vattulainen I, Grant M, Microscopic mechanism for cold denaturation, Phys Rev Lett 2008;100:118101.
 
\bibitem{bennaim2} Ben-Naim A J, Statistical Mechanical Study of Hydrophobic Interaction. I. Interaction between Two Identical Nonpolar Solute Particles, Chem Phys 1971;54:1387
.
\bibitem{yoshidome} Yoshidome T, Harano Y, Kinoshita M, Hydrophobicity at low temperatures and cold denaturation of a protein, Phys Rev E 2009;79:011912.

\bibitem{Kusalik} Kusalik P G, Patey G N, The solution of the reference hypernetted-chain approximation for water-like models, Mol Phys 1988;65:1105-1119.

\bibitem{marques} Marqués M I, Borreguero J M, Stanley H E, Dokholyan N V, Possible Mechanism for Cold Denaturation of Proteins at High Pressure, Phys Rev Lett 2003;91:138103.

\bibitem{sastry} Sastry S, Debenedetti P G, Sciortino F, Stanley H E, Singularity-free interpretation of the thermodynamics of supercooled water, Phys Rev E 1996;53:6144-6154.

\bibitem{zhang} Zhang J, Peng X, Jonas A, Jonas J, NMR study of the cold, heat, and pressure unfolding of ribonuclease A, Biochemistry 1995;34:8631-8641.

\bibitem{lassalle} Lassalle M W, Yamada H, Akasaka K,  The pressure-temperature free energy-landscape of staphylococcal nuclease monitored by 1H NMR, J Mol Biol 2000;298:293-302.

\bibitem{patel} Patel B A, Debenedetti P G, Stillinger F H, Rossky P J, A water-explicit lattice model of heat-, cold-, and pressure-induced protein unfolding, Biophys J 2007;93:4116-4127.

\bibitem{franzese6} Franzese G, Bianco V, Iskrov S, Water at the interface with proteins, Food Biophys 2010; arXiv:1010.4984v1, in print.

\bibitem{soper} Soper A, Ricci M, Structures of high-density and low-density water, Phys Rev Lett 2000;84:2881-2884.

\bibitem{kumar} Kumar P, Buldyrev S V, Starr F W, Giovanbattista N, Stanley H E, Thermodynamics, structure, and dynamics of water confined between hydrophobic plates, Phys Rev E 2005;72:051503.

\bibitem{franzese} Franzese G, Stanley H E, A theory for discriminating the mechanism responsible for the water density anomaly, Physica A 2002;314:508-513.

\bibitem{franzese2} Franzese G, Stanley H E, Liquid-liquid critical point in a Hamiltonian model for water: analytic solution, J Phys Condens Matter 2002;14:2201-2209.

\bibitem{franzese3} Franzese G, Marqu\'es M I, Stanley H E,  Intramolecular coupling as a mechanism for a liquid-liquid phase transition, Phys Rev E 2003;67:011103.

\bibitem{franzese4} Franzese G, Stanley H E, The Widom line of supercooled water, J Phys Condens Matter 2007;19:205126.

\bibitem{debenedetti} Debenedetti P G, Matastable Liquids: Concepts and Principles, Princeton University Press, Princeton, 1996.

\bibitem{ricci} Ricci M A, Bruni F, Giuliani A, Similarities between confined and supercooled water, Faraday Discuss 2009;141:347-358.

\bibitem{franzese5} Franzese G, Santos F d l, Dynamically Slow Processes in Supercooled Water Confined Between Hydrophobic Plates, J Phys Condens Matter 2009;21:504107.

\bibitem{mazza} Mazza M G, Stokely K, Strekalova E G, Stanley H E, Franzese G, Cluster Monte Carlo and numerical mean field analysis for the water liquid-liquid phase transition, Comp Phys Comm 2009;180:497-592.

\bibitem{stokely} Stokely K, Mazza M G, Stanley H E, Franzese G, Effect of hydrogen bond cooperativity on the behavior of water, Proc Natl Acc Sci USA 2010;107:1301-1306.

\bibitem{dlsf} Santos F d l, Franzese G in preparation.

\bibitem{kumar2} Kumar P, Franzese G, Stanley H E, Dynamics and thermodynamics of water, J Phys Condens Matter 2008;20:244114.

\bibitem{kumar3} Kumar P, Franzese G, Stanley H E, Predictions of Dynamic Behavior under Pressure for Two Scenarios to Explain Water Anomalies, Phys Rev Lett 2008;100:105701.

\bibitem{franzese9} Franzese G, Mart\'inez A H, Kumar P, Mazza M G, Stokely K, Strekalova E G, Santos F d l, Stanley H E, Phase transitions and dynamics of bulk and interfacial water, J Phys Condens Matter 2010;22:284103.

\bibitem{chu}  Chu X -q, Faraone A, Kim C, Fratini E, Baglioni P, Leao J B, Chen S -H, Proteins Remain Soft at Lower Temperatures under Pressure, J Phys Chem B 2009;113:5001-5006.

\bibitem{settles} Settles M, Doster W, Anomalous diffusion of adsorbed water: A netron scattering study of hydrated myoglobin, Faraday Discuss 1996;103:269-279.

\bibitem{doster} Doster W, The protein-solvent glass transition, Biochimica et Biophysica Acta 2010;1804:3-14.

\bibitem{speedy} Speedy R J, Limiting forms of the thermodynamic divergences at the conjectured stability limits in superheated and supercooled water, J Phys Chem 1982;86:3002-3005.

\bibitem{debenedetti2} Debenedetti P G, Supercooled and glassy water, J Phys Condens Matter 2003;15:R1669–R1726.

\bibitem{poole} Poole P, Sciortino F, Essmann U, Stanley H E, Phase-Behavior Of Metastable Water, Nature 1992;360:324-328.

\bibitem{poole2} Poole P H, Sciortino F, Grande T, Stanley H E, Angell C A, Effect of Hydrogen Bonds on the Thermodynamic Behavior of Liquid Water, Phys Rev Lett 1994;73:1632–1635.

\bibitem{tanaka} Tanaka H, A self-consistent phase diagram for supercooled water, Nature 1996;380:328–330.

\bibitem{tanaka2} Tanaka H, Phase behaviors of supercooled water: Reconciling a critical point of amorphous ices with spinodal instability, J Chem Phys 1996;105:5099–5111.

\bibitem{angell} Angell C A,  Insights into Phases of Liquid Water from Study of Its Unusual Glass-Forming Properties, Science 2008;319:582-587.

\bibitem{xu} Xu L, Kumar P, Buldyrev S V, Chen S -H, Poole P H, Sciortino F, Stanley H E, Relation between the Widom line and the dynamic crossover in systems with a liquid-liquid phase transition, Proc Natl Acad Sci USA 2005;102:16558-16562.

\bibitem{mazza2} Mazza M G, Stokely K, Pagnotta S E, Bruni F, Stanley H E, Franzese G, Two dynamic crossovers in protein hydration water and their thermodynamic interpretation, arXiv:0907.1810v1, 2009.

\bibitem{mazza3} Mazza M G, Stokely K, Stanley H E, Franzese G, Anomalous specific heat of supercooled water, arXiv:0807.4267, 2008.

\bibitem{coniglio} Coniglio A, Klein W, Cluster and Ising critical droplets: a renormalization group approach, J Phys A 1980;12:2775-2780.

\bibitem{fortuin} Fortuin C M, Kasteleyn P W, On the random cluster model. I. Introduction and relation to other models, Physica (Amsterdam) 1972;57:536-564.

\bibitem{swendsen} Swendsen R H, Wang J S, Nonuniversal critical dynamics in Monte Carlo simulations, Phys Rev Lett 1987;58:86-88.

\bibitem{wolff} Wolff U, Collective Monte Carlo Updating for Spin Systems, Phys Rev Lett 1989;62:361-364.

\bibitem{coniglio2} Coniglio A, Liberto F d, Monroy G, Peruggi F, Exact relations between droplets and thermal fluctuations in external field, J Phys A 1989;22:L837-L842.

\bibitem{cataudella} Cataudella V, Franzese G, Nicodemi M, Scala A, Coniglio A, Critical clusters and efficient dynamics for frustrated spin models, Phys Rev Lett 1994;72:1541–1544. 

\bibitem{franzese10} Franzese G, Cluster analysis for percolation on a two-dimensional fully frustrated system, J Phys A 1996;29:7367-7375.

\bibitem{franzese11} Franzese G,  Potts fully frustrated model: Thermodynamics, percolation, and dynamics in two dimensions, Phys Rev E 2000;61:6383-6391.

\bibitem{franzese12} Franzese G, Cataudella V, Coniglio A, Invaded cluster dynamics for frustrated models, Phys Rev E 1998;57:88–93.

\bibitem{bianco} Bianco V, Franzese G, in preparation.

\bibitem{franzese8} Franzese G, Stokely K, Chu X -q, Kumar P, Mazza M G, Chen S -H, Stanley H E,  Pressure effects in supercooled water: comparison between a 2D model of water and experiments for surface water on a protein, J Phys Condens Matter 2008;20:494210.

\bibitem{strekalova} Strekalova E G, Mazza M G, Stanley H E, Franzese G, Large decrease of fluctuations for supercooled water in hydrophobic nanoconfinement, arXiv:1010.0693v1, 2010.

\bibitem{Vendruscolo}
We thanks M. Vendruscolo for discussing this point. 

\bibitem{stillinger} Stillinger F H, Structure in aqueous solutions of nonpolar solutes from the standpoint of scaled-particle theory, J Sol Chem 1973;2:141-158.

\bibitem{franzese7} Franzese G, Coniglio A, Phase transitions in the Potts spin-glass model, Phys Rev E 1998;58:2753-2759.

\bibitem{tenwolde} tenWolde P R, Frenkel D, Enhancement of protein crystal nucleation by critical density fluctuations, Science 1997;277:1975-1978.

\bibitem{franzese13} Franzese G, Malescio G, Skibinsky A, Buldyrev S V, Stanley H E, Metastable liquid-liquid phase transition in a single-component system with only one crystal phase and no density anomaly, Phys Rev E 2002;66:051206.

\bibitem{stokely2} Stokely K, Mazza M, Stanley H E, Franzese G, Metastable Systems under Pressure, Eds: Rzoska S J, Drozd-Rzoska A, Mazur V A; Springer; pp 197-216, 2010. 

\end{thebibliography}
\end{document}